\documentclass[prl,twocolumn,showpacs]{revtex4-1}
\usepackage[dvipdfmx]{graphicx}
\usepackage{amsmath,amssymb}
\usepackage{amssymb}   % for math
\usepackage{bm}
\usepackage{color}
\usepackage[colorlinks=true,linkcolor=blue]{hyperref}%
\begin{document}
\title{New chaos indicators for systems with extremely small Lyapunov exponents}%
\author{Ken-ichi Okubo}%
\email{okubo.kenichi.65z@st.kyoto-u.ac.jp}
\author{Ken Umeno} %
\affiliation{Department of Applied Mathematics and Physics, 
Graduate School of Informatics, Kyoto University} %
\date{January 2015}%
%\tableofcontents
\begin{abstract}
We propose new chaos indicators for systems with extremely small positive Lyapunov exponents. These chaos indicators can firstly detect a sharp transition between the Arnold diffusion regime and the Chirikov diffusion regime of the Froeschl\'e map and secondly detect chaoticity in systems with zero Lyapunov exponent such as the Boole transformation and the $S$-unimodal function to characterize sub-exponential diffusions. 
\end{abstract}
\maketitle

\paragraph{Introduction}
 In weakly  chaotic systems with extremely  small Lyapunov exponents, it is well-known that it takes a very long time to estimate the largest Lyapunov exponent in the order of which is inversely propositional of the largest Lyapunov exponent. 
Thus, there is a practice that it can take more than ten times longer than the Lyapunov time \cite{Froeschle97a}. 
For example, G. J. Sussman and J. Wisdom \cite{Sussman92} numerically showed the Lyapunov time of the solar system is approximately $5\times 10^6$ years
by computing over $1\times 10^8$ years. 
For investigating nearly integrable systems with such weak chaotic property, Froeschl\'e et al. proposed 
a chaos indicator called {\it Fast Lyapunov Indicator} (FLI) \cite{Froeschle97a,Froeschle97b}. If an initial point belongs to a chaotic domain, the time evolution of FLI grows linearly. On the contrary,
if an initial point belongs to a torus domain, the time evolution of FLI grows logarithmically \cite{Guzzo}. Besides the fact that the original key concept of FLI has not been changed, OFLI \cite{Fouchard} and
OFLI$^2$ \cite{Barrio05} are proposed as the improvements of FLI, which can reduce the dependency of direction of initial variational vectors. 
In addition to nearly integrable systems, in infinite ergodic  systems with zero  Lyapunov exponents, the sub-exponential behavior
attracts lots of interests \cite{Akimono,Akimoto15}. Akimoto et al. \cite{Akimoto15} proposed generalized Lyapunov exponent which characterizes super-exponential chaos and sub-exponential chaos. 
That is because the finite time Lyapunov exponent converges in distribution \cite{Aaronson97,Akimono}. If an orbital expansion rate $\Delta$ grows in order of 
$\Delta \sim \exp\left( n^\alpha\right), 0<\alpha<1$, we need to get index $\alpha$ to applying the general Lyapunov exponent.

In this Communication, we propose a new chaos indicator that can detect chaoticity of weak chaotic systems with extremely small positive Lyapunov exponent more rapidly than
these existing methods FLI, OFLI and OFLI$^2$. 
In addition, this new chaos indicator can firstly detect a sharp transition between Arnold diffusion and Chirikov diffusion. 
Then, we propose another new indicator which can find the index $\alpha$ about sub-exponential chaos whose orbital expansion rate $\Delta$ grows 
$\Delta \sim \exp(n^\alpha), 0<\alpha<1$
occuring in  the Boole transformation \cite{Adler,Akimono} and the $S$-unimodal function \cite{Thaler83}.
\paragraph{Ultra Fast Lyapunov Indicator}
We assume such a dynamical system as
\begin{eqnarray}
\bm{f} &:& \mathbb{R}^N \to \mathbb{R}^N,\nonumber \\
\bm{x}_{n+1} &=& \bm{f}(\bm{x}_n).
\end{eqnarray}
and let us consider such a variation $\bm{\delta}$ as
\begin{eqnarray*}
\bm{x}_{n+1} &=& \bm{x}_n + \bm{\delta}_n,\\
&=& \bm{f}(\bm{x}_n),\\
&=& \bm{f}(\bm{x}_{n-1} + \bm{\delta}_{n-1}),\\
&=& \bm{f}(\bm{x}_{n-1}) + D\bm{f}(\bm{x}_{n-1})\bm{\delta}_{n-1} + \frac{1}{2}D^2\bm{f}(\bm{x}_{n-1})(\bm{\delta}_{n-1})^2\\
& &  + O(\bm{\delta}_{n-1}^3),
\end{eqnarray*}
then, we get
\begin{eqnarray}
\begin{split}
\bm{\delta}_n =& D\bm{f}(\bm{x}_{n-1})\bm{\delta}_{n-1} +\frac{1}{2}D^2\bm{f}(\bm{x}_{n-1})(\bm{\delta}_{n-1})^2&\\ 
&+ O(\bm{\delta}_{n-1}^3),&
 \end{split}
\end{eqnarray}
where 
\begin{eqnarray*}
D^2\bm{f}(\bm{x}_{n})(\bm{\delta}_{n})^2 \equiv \sum_{i=1}^{N}{}^t\!\bm{\delta}_n \mathcal{H}[f_i(\bm{x}_n))] \bm{\delta}_n \bm{e}_i,
\end{eqnarray*}
where $\mathcal{H}$ is a Hessian matrix, $f_i(\bm{x}_n)$ is an $i$th component of $\bm{f}(\bm{x}_n)$ and $\bm{e}_i$ is a unit vector about $i$th component.
Usually, we ignore the terms whose order is greater than one and use such a variational equation as
\begin{eqnarray}
\bm{\delta}_{n+1} = D\bm{f}(\bm{x}_n)\bm{\delta}_n \label{first order variational equation}.
\end{eqnarray}
We apply Eq.(\ref{first order variational equation}) to compute the largest Lyapunov exponent and FLI.
We propose a new indicator called {\it Ultra Fast Lyapunov Indicator (UFLI)} with second order derivatives 
in order to detect chaoticity more rapidly and clearly as follows.
The definition of UFLI is 
\begin{eqnarray}
\mbox{UFLI}(\bm{x}_0,\bm{w}_0,T) \equiv \sup_{0<i \leq T} \log\left(\frac{\|\bm{w}_i^\perp\|}{\|\bm{w}_0\|} \right), \\
\bm{w}_{n+1}= D\bm{f}(\bm{x}_n)\bm{w}_n + \frac{1}{2}D^2\bm{f}(\bm{x}_n)\left( \bm{w}_n\right)^2, \label{variational equation}\\
\|\bm{w}_n^\perp\| = \sqrt{\|\bm{w}_n\|^2 -\frac{\left\langle \bm{w}_n, \bm{f}(\bm{x}_n) \right\rangle ^2}{\|\bm{f}(\bm{x}_n)\|^2}}, 
\end{eqnarray}
 where $\bm{w}_n, D\bm{f}(\bm{x}_n), D^2\bm{f}(\bm{x}_n)\left( \bm{w}_n\right)^2, \bm{w}_n^\perp$ are a variational vetor, a Jacobian of $\bm{f}(\bm{x}_n)$, a vector whose $i$th component consists of a product between ${}^t\!\bm{w}_n$, Hessian matrix $\mathcal{H}[f_i(\bm{x}_n))]$ and $\bm{w}_n$ where $f_i(\bm{x}_n)$ is an $i$th component of $\bm{f}(\bm{x}_n)$ and a orthogonal component of  $\bm{w}_n$ respectively. This proposal is different from the work by Dressler, Farmer \cite{Farmer} and Taylor \cite{Taylor} who introduce generalized Lyapunov exponents using higher derivatives and the work by Barrio \cite{Barrio05}.
 
\begin{center}
\begin{figure}
\includegraphics[width =8cm]{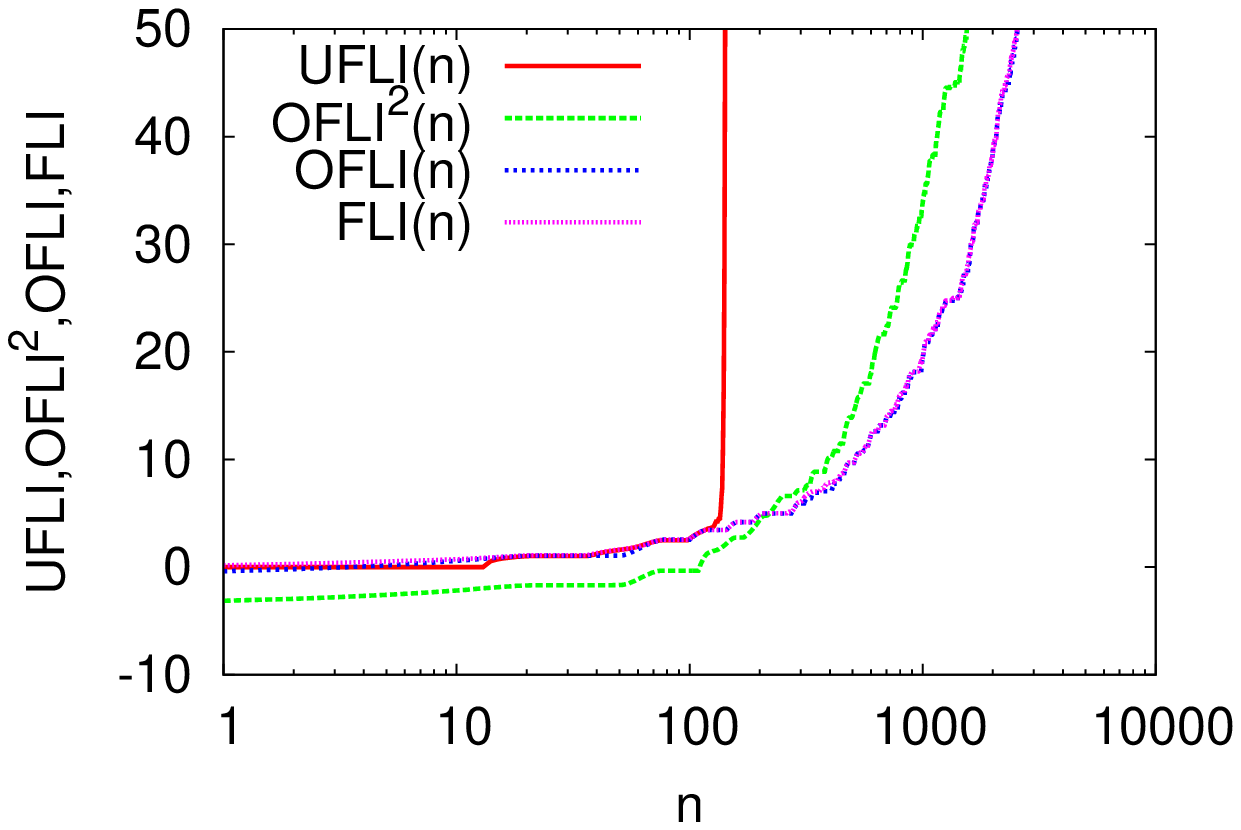}
\vspace{5mm}
\caption{Point A. A time evolution of UFLI, OFLI$^2$, OFLI and FLI in a chaos domain. The common logarithm is used to calculate
UFLI, OFLI$^2$, OFLI and FLI.}
\label{PointA}
\includegraphics[width = 8cm]{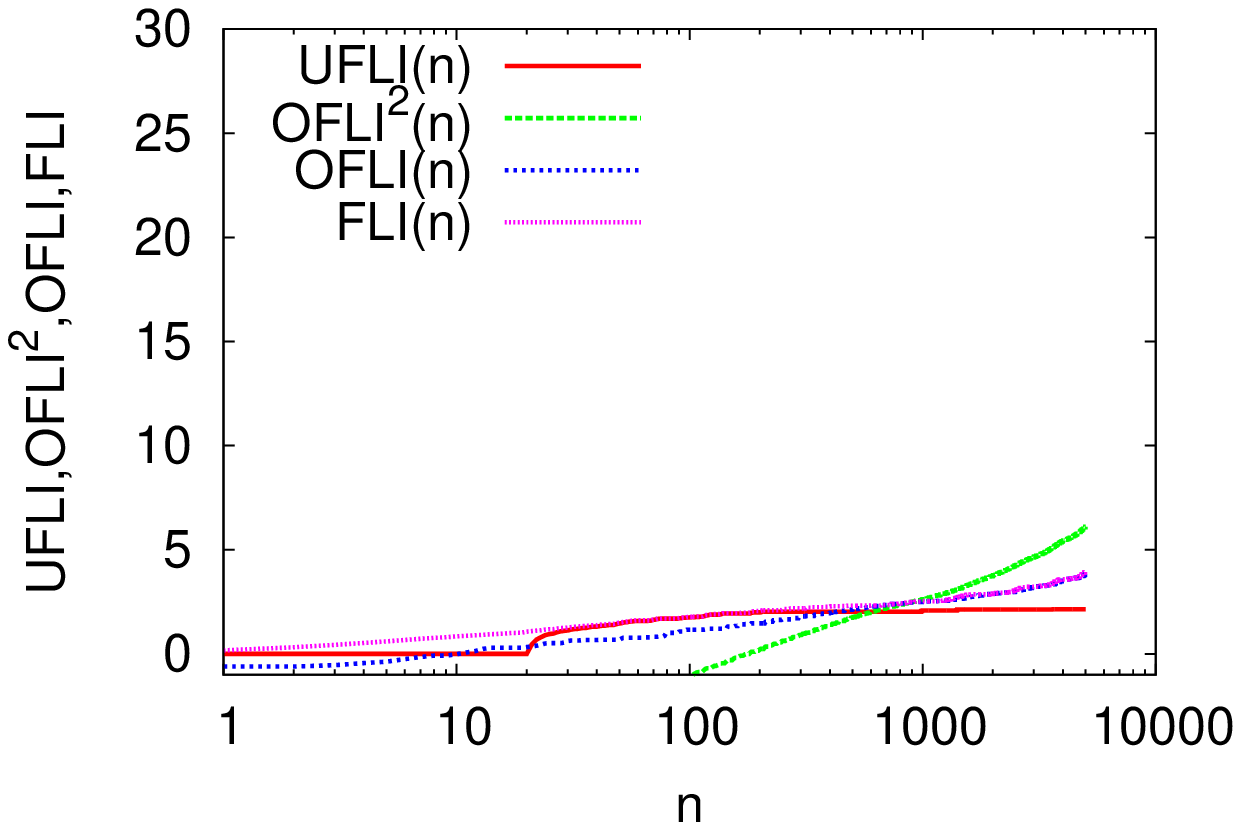}
\vspace{5mm}
\caption{Point B. A time evolution of UFLI, OFLI$^2$, OFLI and FLI in a KAM torus domain. The common logarithm is used to calculate
UFLI, OFLI$^2$, OFLI and FLI.}
\label{PointB}
\includegraphics[width = 8cm]{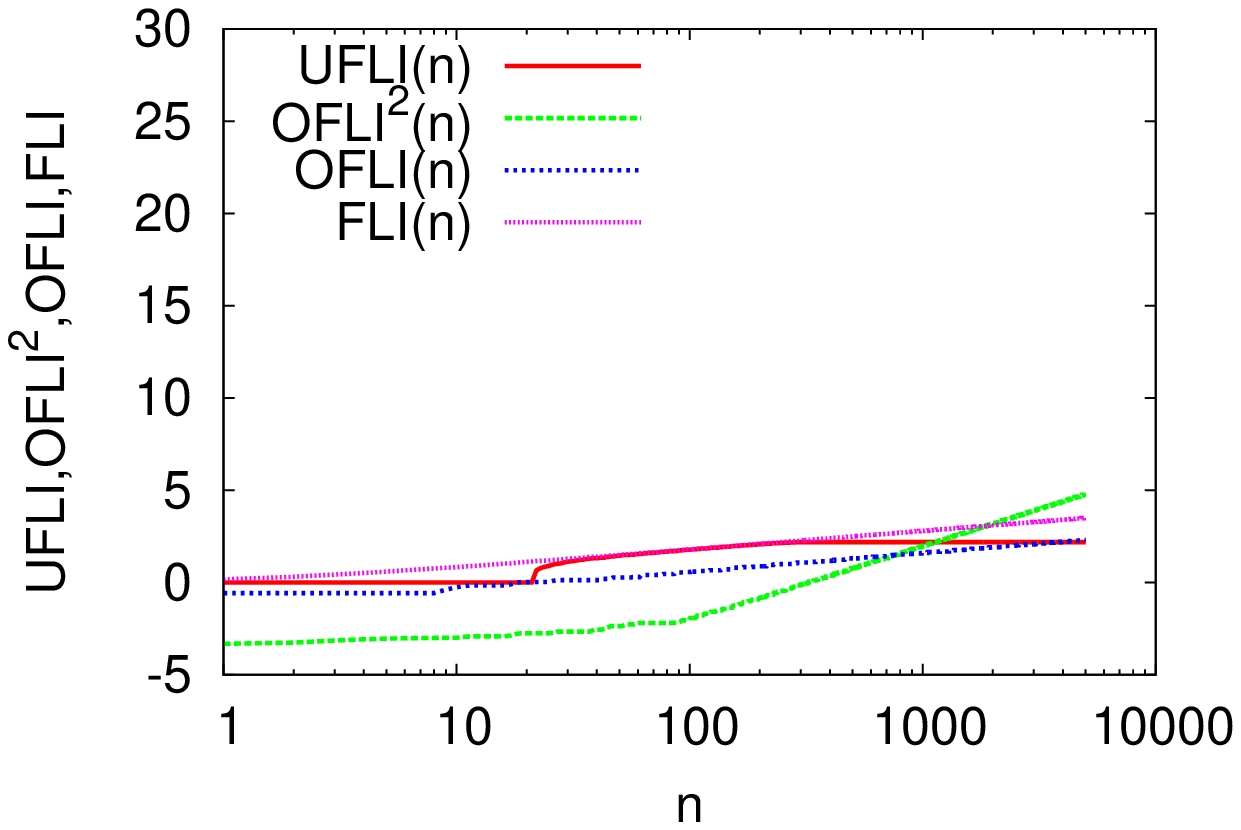}
\vspace{5mm}
\caption{Point C. A time evolution of UFLI, OFLI$^2$, OFLI and FLI in a resonant torus domain. The common logarithm is used to calculate
UFLI, OFLI$^2$, OFLI and FLI.}
\label{PointC}
\end{figure}
\end{center}

%(ここでしっかりと定義しないといけない。)
The formula (\ref{variational equation}) shows a variational equation considering a second order derivative. We explain the ability of UFLI.
Let us define $\bm{u}$ and inner product $B_i^n$ by
\begin{eqnarray}
\bm{u}_n &\equiv& D\bm{f}_{n-1}D\bm{f}_{n-2}\cdots D\bm{f}_0 \bm{w}_0,\\
B_i^n &\equiv& {}^t\!\bm{u}_n \mathcal{H}[f_i(\bm{x}_n))]\bm{u}_n,\\
B_i^n \bm{e}_i &\equiv& \sum_{i=1}^{N} B_i^n\bm{e}_i.
\end{eqnarray}
Then, $\bm{v}_{n+1}$ is expressed by
\begin{widetext}
\begin{eqnarray}
\bm{w}_{n+1} &=& D\bm{f}_n\bm{u}_n + \frac{1}{2}\sum_{l=0}^n\left(\prod_{j=1}^l D\bm{f}_{j}\right) B_i^{n-l}\bm{e}_i + O(|\bm{w}_0|^3),\\
&=& \left(\prod_{i=0}^{n}D\bm{f}_i\right) \bm{w}_0+ \frac{1}{2}\sum_{l=0}^n\left(\prod_{j=1}^l D\bm{f}_{j}\right) B_i^{n-l}\bm{e}_i + O(|\bm{w}_0|^3). \label{variational equation1}
\end{eqnarray}
\end{widetext}
For example, at $l=n$, $\left(\prod_{j=1}^n D\bm{f}_{j}\right)B_i^{0}\bm{e}_i $ is considered as a variational vector evolved by 
Eq. (\ref{first order variational equation}) whose initial vector is $B_i^{0}\bm{e}_i$. Then, $\left(\prod_{j=1}^n D\bm{f}_{j}\right)B_i^{0}\bm{e}_i $ 
expands in the direction of the Largest Lyapunov exponent. In the same way, higher order terms are considered as variational vectors evolved by 
Eq. (\ref{first order variational equation}) if $n$ is enough large. Therefore, $\|\bm{w}_n\|$ grows drastically.
The time evolution of UFLI changes clearly if an initial point belongs to chaos domain and grows slowly if the initial point belongs to a domain of KAM- or Resonant torus.
Here, We apply UFLI to the Froeschl\'e map which is known to show Arnold diffusion and Chirikov diffusion \cite{Froeschle00,Froeschle05,Froeschle06}, where the map is defined by 
\begin{eqnarray}
T_F
\left(
\begin{array}{l}
I_1\\
\theta_1\\
I_2\\
\theta_2
\end{array}\right)
&=&\left(
    \begin{array}{l}
      I_1-\varepsilon\frac{\sin(I_1+\theta_1)}{\{\cos(I_1+\theta_1)+\cos(I_2+\theta_2)+4\}^2} \\
      I_1+\theta_1  \pmod{2\pi}\\
      I_2-\varepsilon\frac{\sin(I_2+\theta_2)}{\{\cos(I_1+\theta_1)+\cos(I_2+\theta_2)+4\}^2}\\
      I_2+\theta_2  \pmod{2\pi}
    \end{array}
  \right) \label{Froechlemap2},
\end{eqnarray}
where $I_{1},I_{2}$ are action variables and $\theta_1, \theta_2$ are action-angle variables correspondng to action variables respectively. %・・・という様に ここでも変数の説明が必要
Figures. \ref{PointA}, \ref{PointB} and \ref{PointC} show the time evolutions of UFLI, OFLI$^2$, OFLI and FLI with the initial points A, B and C respectively.
The float128 precision is used to calculate them.
Three initial points A$=(I_1,\theta_1,I_2,\theta_2)=(2.04, 0, 2.1, 0)$, B$=(1.8, 0, 1.2, 0)$, 
C$=(1.67,0,0.91,0)$ correspond to the chaotic domain, the KAM torus domain and the resonant torus domain respectively in the Froeschl\'e map with $\varepsilon=0.6$
 \cite{Froeschle05}. 
We set the initial variational vector as below.
\begin{eqnarray}
\left\lbrace 
\begin{array}{lll}
w_1(0) &=& 0.001,\\
w_2(0) &=& 0.001,\\
w_3(0) &=& \frac{\sqrt{3}-1}{2}\times 0.001,\\
w_4(0) &=& 0.001,\\
\|w(0)\| &=& \frac{\sqrt{16-2\sqrt{3}}}{2}\times 0.001~\sim 0.0017.
\end{array}\right.
\label{initialcondition}
\end{eqnarray}

\begin{center}
\begin{figure}[t]
\vspace*{-25pt}
\includegraphics[width = 8cm]{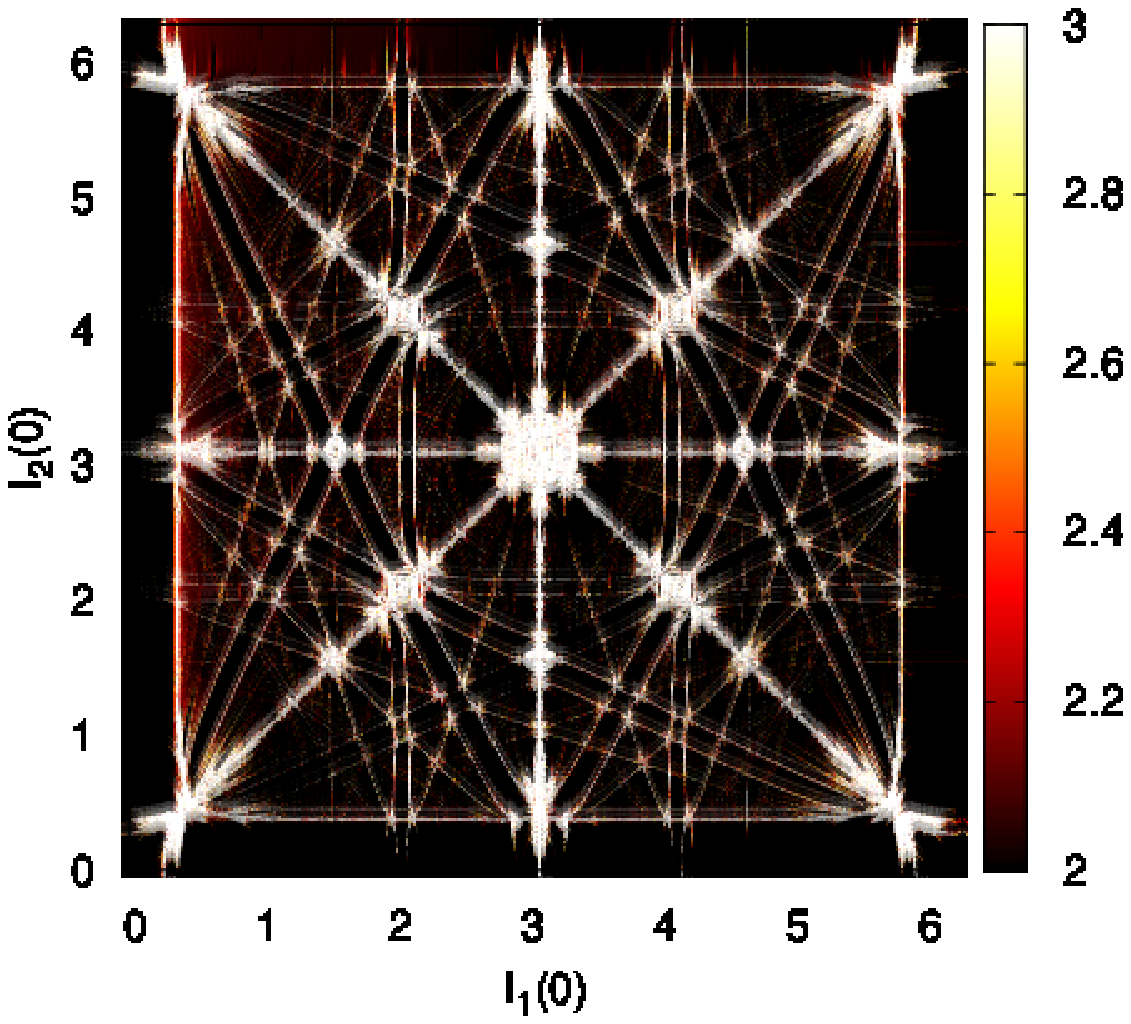}
\vspace{-25pt}
\caption{A digaram of UFLI for Froeschl\'e map with $\varepsilon=0.6$ at $n=200$. The common logarithm is used to calculate
UFLI.}
\label{UFLIdiagram}
\includegraphics[width = 8cm]{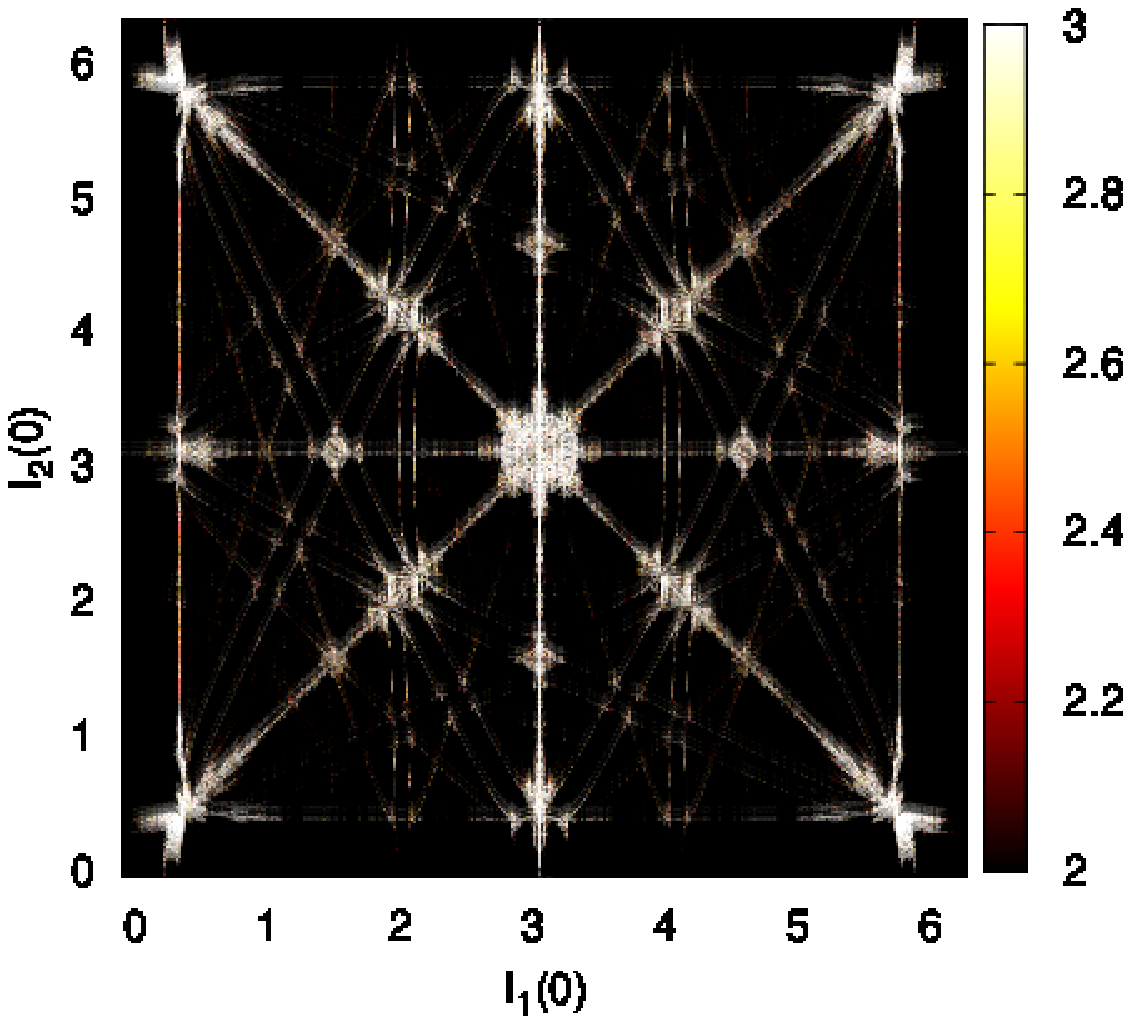}
\vspace{-25pt}
\caption{A diagram of OFLI$^2$ for Froeschl\'e map with $\varepsilon=0.6$ at $n=200$. The common logarithm is used to calculate
OFLI$^2$.}
\label{OFLI2diagram}
\end{figure}
\end{center}

\begin{center}
\begin{figure}[t]
\includegraphics[width=8cm]{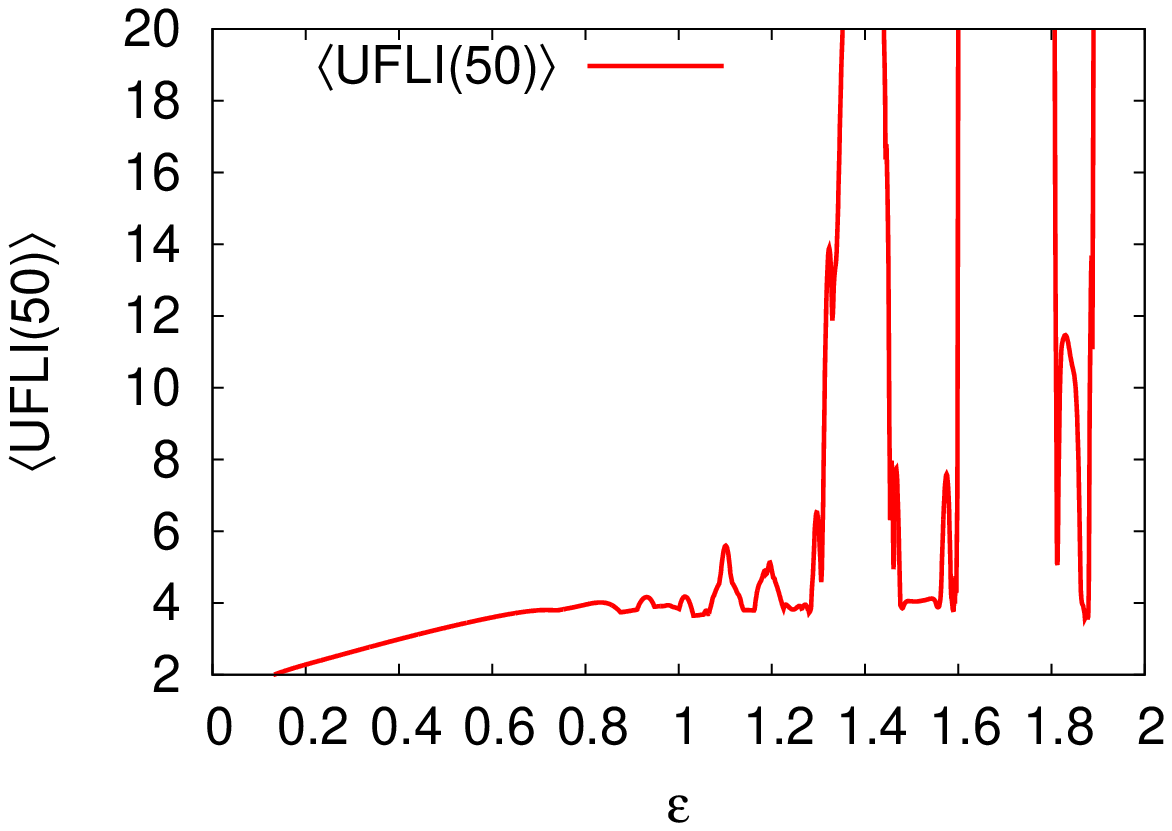}
\vspace{5mm}
\caption{The ensemble average of UFLI$(50)$ v.s. $\varepsilon$. The common logarithm is used to calculate
UFLI.}
\label{UFLI for each epsilon}
\includegraphics[width=8cm]{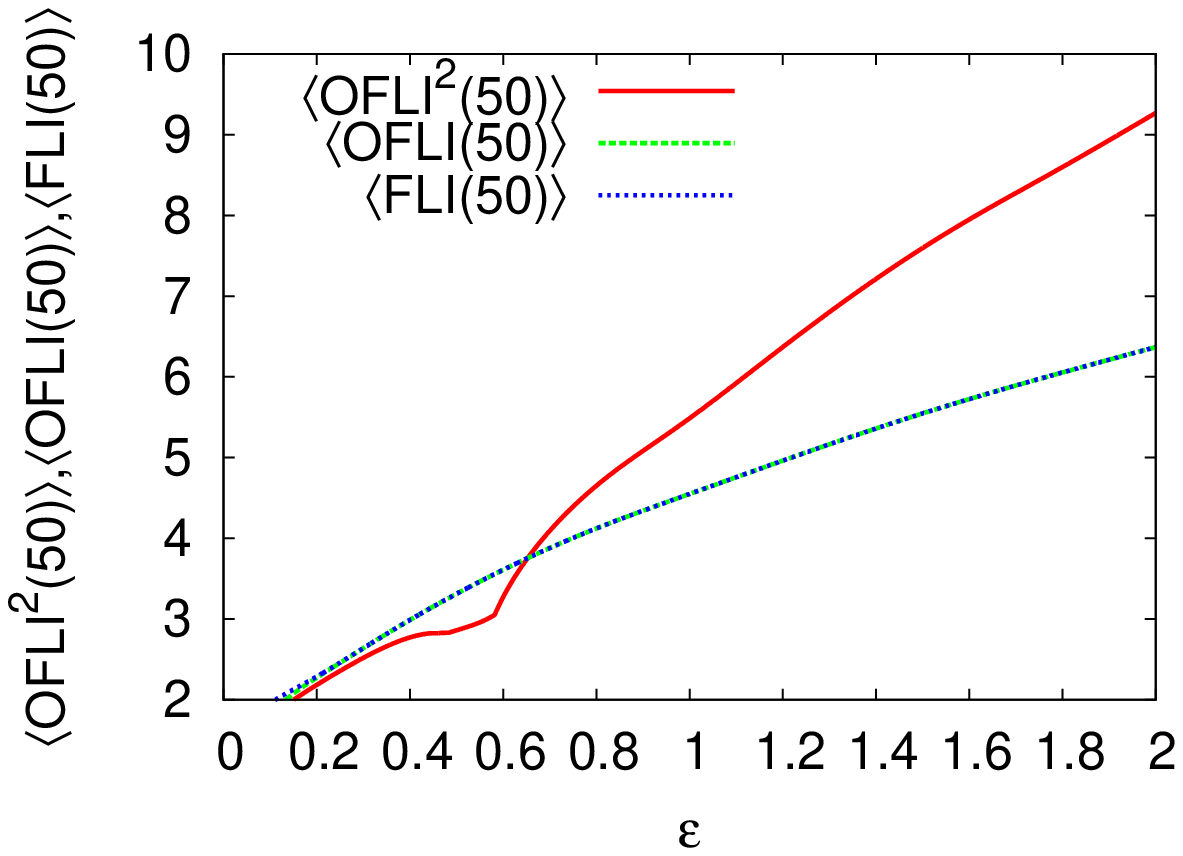}
\vspace{5mm}
\caption{The ensemble average of OFLI$^2(50)$, OFLI(50) and FLI(50) v.s. $\varepsilon$. The common logarithm is used to calculate
OFLI$^2$.}
\label{OFLI2 for each epsilon}
\end{figure}
\end{center}

According to Figs. \ref{PointA}, \ref{PointB} and \ref{PointC}, our proposed UFLI performs much better compared to OFLI$^2$, OFLI and FLI.
Figures \ref{UFLIdiagram} and \ref{OFLI2diagram} show diagrams of UFLI and OFLI$^2$ for Froeschl\'e map with $\varepsilon=0.6$ at $n=200$ whose 
initial condition is $\theta_1=\theta_2=0$. UFLI can show Arnold web, a structure consists of resonant lines more clearly than OFLI$^2$.
According to Ref. \cite{Froeschle06}, this map behaves differently as a magnitude of $\varepsilon$. It is known in Ref. \cite{Froeschle06} that , Arnold diffusion occurs when
$\varepsilon \leq 0.9$ and Chirikov diffusion occurs when $\varepsilon \geq 1.3$. Here, we apply UFLI to detect a change between these  diffusion regime. 
We compare variations of UFLI and OFLI$^2$ v.s. $\varepsilon$.
One thousand initial points are chosen near $(I_1,I_2)=(\pi/2,\pi/2)$. 
Figure \ref{UFLI for each epsilon} shows 
ensemble average of UFLI$(50)$ v.s. the parameter $\varepsilon$ change and Fig. \ref{OFLI2 for each epsilon} shows the counterpart of OFLI$^2(50)$, OFLI(50) and FLI(50) instead of UFLI$(50)$.
Figure \ref{UFLI for each epsilon} shows that UFLI loses smoothness in $\varepsilon \geq 0.9$ and 
distinguishes a transition between the two regimes (Arnold diffusion and Chirikov diffusion) of Froeschl\'e map although OFLI$^2$ and the other existing detectors such as FLI cannot detect any transition in Fig. \ref{OFLI2 for each epsilon}. We define Lyapunov time $T_L$ and the macroscopic instability time $T_I$.
In Arnold diffusion, $T_I \sim \exp\left( T_L\right)$ while  in Chirikov diffusion, $T_I$ is order of polynomial for $T_L$ \cite{Morbidelli}. 
Then, it is important to detect the point of $\varepsilon$ around which diffusion type changes.

\begin{center}
\begin{figure}[t]
%\hspace*{-0.5cm}
\includegraphics[width = 8cm]{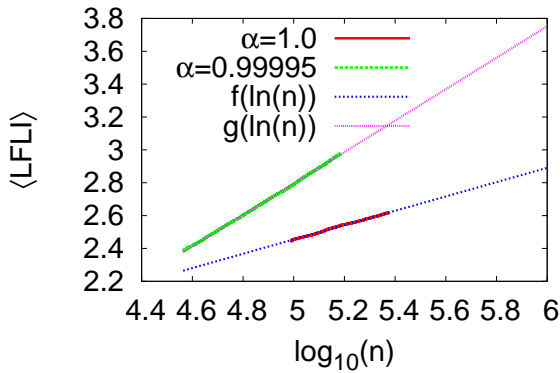}
%\vspace{-5mm}
\vspace*{2mm}
\caption{The ensemble average of the time evolution of LFLI of Boole transformation and generalized Boole transformation. 
The common logarithm is used to calculate LFLI.}
\label{boolean}
\end{figure}
\end{center}

According to the result above, our proposed UFLI chaos detector is very powerful to detect chaoticity of systems with relatively small Lyapunov exponents more rapidly and clearly than FLI, OFLI and OFLI$^2$. In addition to this, 
UFLI can also detect a sharp change of the diffusion regime between Arnold diffusion and Chirikov diffusion although OFLI$^2$ and the other existing indicators cannot detect any transition.
\paragraph{Log Fast Lyapunov Indicator}
Here, we investigate further to {\it chaotic} systems 
 with zero Lyapunov exponent.
In generally, a positive Lyapunov exponent shows a existence of exponential growth of a variation  between two close orbits. A positive value of Lyapunov exponent is used as an indicator of 
chaoticity. Even though the value of Lyapunov exponent is zero, behavior on torus and sub-exponential behavior with zero Lyapunov exponent are different. Thus, we propose another new indicator to distinguish them. In this section,
\textit{Log Fast Lyapunov Indicator} (LFLI)
\begin{eqnarray}
\mbox{LFLI}(n) \equiv \log \left[ \sup_{0<i\leq n} \log \left( \frac{\left\|D\bm{f}(\bm{x}_i) \bm{w}_i\right\|}{\left\|\bm{w}_0\right\|}\right)  \right] 
\end{eqnarray}
is proposed to characterize sub-exponential behaviors. Here, $\bm{w}_0, \bm{w}_i, D\bm{f}(\bm{x}_i)$ are an initial variational vetor, a variational vector at $n=i$ and a Jacobian of $\bm{f}(\bm{x}_i)$ respectively. %(等を定義。)
If infinite ergodic systems behave sub-exponentially \cite{Akimono}, the time evolution of LFLI grows linearly with slope smaller than unity. If systems have a positive Lyapunov exponent, the slope is unity.
We apply LFLI to the Boole transformation and the $S$-unimodal function in the following section.
\paragraph{Boole transformation}
Here, the Boole transformation $T:\mathbb{R}\to \mathbb{R}$ is defined by 
\begin{eqnarray}
x_{n+1} =T(x_n) \equiv x_n -\frac{1}{x_n}.
\end{eqnarray}
It is known that the Boole transformation is ergodic and preserves the Lebesgue measure \cite{Adler}. The Boole transformation is an infinite ergodic system and the following equation is known to hold %[Here $m$ is a 何の？measure.ここの部分を式の後に]
\begin{eqnarray}
\lim_{n\to\infty}\frac{1}{n}\sum_{k=1}^n f(T^kx) =0, \\
\mbox{  a.e. } x \in \mathbb{R},~{}^\forall\!f \in L^1(\mu), \nonumber
\end{eqnarray}
where $\mu$ is the invariant measure for the probability preserving Boole transformation $T$ \cite{Aaronson97}. 
By substituting $f(T^kx)=\log|T'(x_k)|$, we know that the a value of Lyapunov exponent of Boole transformation is zero. However, 
it is known that the dynamical system behaves sub-exponentially \cite{Akimono}. Namely,  its orbital expansion rate $\Delta$ grows $\Delta \sim \exp(t^\frac{1}{2})$.
By using the LFLI, we can find a power index. To compare with the Boole transformation, we consider the following generalized Boole transformations 
\begin{eqnarray}
x_{n+1} =T_{\alpha, ~\beta}(x_n) \equiv \alpha x_n -\frac{\beta}{x_n},\\0<\alpha<1,0<\beta,\nonumber
\end{eqnarray}
which are known to have non-negative Lyapunov exponent \cite{Umeno}
\begin{eqnarray}
\lambda = \log\left( 1+2\sqrt{\alpha(1-\alpha)}\right).
\end{eqnarray}
We put $\beta=\alpha$ here for simplicity, because $\beta$ doesn't affect on the Lyapunov exponent $\lambda$.
Figure \ref{boolean} shows ensemble averages of the time evolution of LFLI for the Boole transformation and the generalized Boole transformations with $\alpha=0.99995$ whose 
three hundred initial points are chosen near a point $x=11.7$.
Here, the initial condition is $\bm{w}(0)=0.00000000000000000001$ and the float 128 precision is used to calculate LFLI.

In Fig. \ref{boolean}, the  $f(\ln(n))$ and $g(\ln(n))$ are linear approximations of the ensemble averages of
the Boole transformation and the generalized Boole transformations respectively.
The slopes of $f(\ln(n))$ and $g(\ln(n))$ are  about $0.436$ and $0.957$ respectively. These results indicate that our proposed LFLI is very powerful  to find a power index for sub-exponential behavior.
\paragraph{$S$-unimodal function}
$S$-unimodal function \cite{Thaler83} is defined by
\begin{eqnarray}
X_{n+1} = T_S(X_n) = \frac{1-5X_n^2}{1+3X_n^2},\ X_n\in [-1,1].
\end{eqnarray}
$S$-unimodal function has an infinite measure whose density function is 
\begin{eqnarray}
\rho(X) = \frac{1}{(1+X)\sqrt{1-X^2}}.
\end{eqnarray}
According to Ref. \cite{Thaler83}, orbit expansion rate of $S$-unimodal function is $\Delta \sim \exp( n^{\frac{1}{2}})$.
\begin{figure}
\begin{center}
\includegraphics[width=8cm]{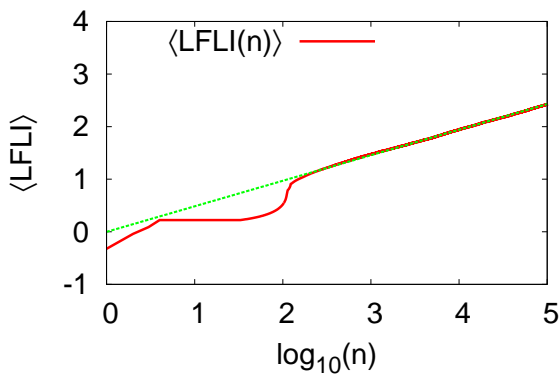}
\vspace*{1mm}
\caption{The time evolution of ensemble average of LFLI of the $S$-unmodal function. A slape of linear approximation for LFLI is about 0.486. 
The common logarithm is used to calculate LFLI.}
\label{S-unmodal function}
\end{center}
\end{figure}
Figure \ref{S-unmodal function} shows the ensemble average of the time evolution of LFLI for the $S$-unmodal function whose
one thousand initial points are chosen near a point $X=e^{-1}$.
Here, the initial condition is $\bm{w}(0) = \exp(-20)$ and the float 128 precision is used to calculate LFLI.

\paragraph{Conclusion}
We propose two chaos indicators Ultra Fast Lyapunov Indicator (UFLI) and Log Fast Lyapunov Indicator (LFLI). It is found that 
UFLI can detect chaoticity more rapidly than OFLI$^2$, OFLI and FLI and the only  UFLI can  detect  a sharp change between Arnold diffusion and Chirikov diffusion regimes, that has not been detected by the existing methods such as OFLI$^2$.
LFLI can measure a power index of a sub-exponential system. In particular, LFLI firstly characterizes 
chaoticity of systems which have zero Lyapunov exponent which has been regarded as non-chaotic systems. Such detectors UFLI and LFLI proposed here are very promising to detect chaoticity of experimental data of intrinsically weakly  chaotic systems.

\end{document}